\documentclass[prb,aps]{revtex4}
\usepackage{graphicx}
\begin{document}

\title{Edge reconstruction in the fractional quantum Hall regime}

\author{Xin Wan}
\affiliation{National High Magnetic Field Laboratory,
Florida State University, Tallahassee, Florida 32310}

\author{E.~H. Rezayi}
\affiliation{Department of Physics, California State University, 
Los Angeles, California 90032}

\author{Kun Yang}
\affiliation{National High Magnetic Field Laboratory and Department
of Physics, Florida State University, Tallahassee, Florida 32306}

\date{\today}

\begin{abstract}

The interplay of electron-electron interaction and confining potential
can lead to the reconstruction of fractional quantum Hall edges.  We
have performed exact diagonalization studies on microscopic models of
fractional quantum Hall liquids, in finite size systems with disk
geometry, and found numerical evidence of edge reconstruction
under rather general conditions. In the present work we have taken into
account effects like layer thickness and Landau level mixing, which are 
found to be of
quantitative importance in edge physics. Due to edge reconstruction,
additional nonchiral edge modes arise for both incompressible and
compressible states. These additional modes couple to electromagnetic fields 
and thus can be detected in microwave conductivity measurements. They are   
also expected to affect the exponent of electron Green's function, which has
been measured in tunneling experiments. We have studied in this work
the electric dipole spectral
function that is directly related to the microwave conductivity
measurement. 
Our results are consistent with the enhanced microwave conductivity
observed in experiments performed on samples with an array of antidots
at low temperatures, and its suppression at
higher temperatures. We also discuss the effects of the edge
reconstruction on the single electron spectral function at the edge.

\end{abstract}


\maketitle


\section{Introduction}

The edge of a quantum Hall (QH) system provides a special environment
to study electron correlations in one dimension. 
Due to the presence of a strong magnetic field and electron-electron
interaction, the bulk of a QH liquid is incompressible, while low-lying 
excitations exist only at the boundary of the liquid. 
In an experimental sample, the physics of edge excitations is 
strongly affected by the interplay of electron-electron interaction 
and the confining potential due to positive background charge.
For example, the edge of an integer QH liquid with a sharp 
confining potential is described by the chiral Fermi-liquid theory
and only a single branch of gapless edge excitations exists
due to the presence of magnetic field.\cite{halperin}
When the confining potential is sufficiently smooth, the edge 
undergoes a reconstruction transition in which a portion 
of the electron liquid is expelled
a few magnetic lengths away from the periphery of the main 
droplet.\cite{macdonald93,chamon94} 
Additional low-lying edge excitations that propagate in {\em both} directions
arise 
after the edge reconstruction transition.~\cite{chamon94} 

The edge excitations of fractional QH liquids are proposed to 
be described by the chiral Luttinger liquid (CLL) 
theory.~\cite{wen92} 
For principal Landau level (LL) filling fractions $\nu = 1/m$, 
the theoretical picture involves only one chiral boson mode. On the other hand 
our exact diagonalization study\cite{wan02} of a microscopic model of 
fractional QH liquids, in finite size systems with disk geometry, 
suggests that a fractional QH liquid 
can undergo edge reconstruction for both smooth {\em and}
sharp confining potentials. 
As a consequence, additional low-energy edge excitations {\em not} described by
the CLL theory are generated and 
introduce complications into the edge physics.
These excitations are clearly visible in the low-energy excitation 
spectrum of a fractional QH system with reconstructed edge.\cite{wan02}

Apart from purely theoretical interest and importance, our study of
fractional quantum Hall edge reconstruction has also been 
motivated by two types of experimental studies of edge physics, both
with puzzling results. 
First, the CLL theory predicts a power-law 
current-voltage dependence ($I \sim V^{\alpha}$)
in the tunneling between a Fermi liquid metal and a QH edge. 
The prediction has been tested by 
experiments~\cite{chang96,grayson98,hilke01,chang01}
using samples made by the cleaved-edge overgrowth technique.~\cite{pfeiffer90}
For a simple filling fraction like $\nu = 1/3$, these
experiments\cite{chang96,grayson98,chang01} 
found non-Ohmic $I$-$V$ dependence with the
exponent $\alpha$ scattering between 2.5 and 2.8, which is
close to but noticeably different from the CLL
theory prediction of a {\em universal} exponent $\alpha = 3$. 
Furthermore, no convincing plateau behavior away from $\nu =
1/3$ is present~\cite{grayson98,chang01} as predicted by the theory. 
In fact, for $1/3 < \nu < 1$, $\alpha$ seems to vary continuously 
for both compressible and incompressible values of $\nu$,
and no universality in $\alpha$ can be extracted from data 
available to date.~\cite{grayson98,hilke01}
Experimental findings have prompted a number of 
theories,~\cite{conti98,han97,zuelicke99,lee98,lopez99,levitov01,goldman,mandal02}
most of which address the continuous dependence of 
$\alpha \sim 1 / \nu$ found in one experiment~\cite{grayson98} only. 

The second type of experiments measure the microwave conductivity 
of a 2DEG with an array of antidots (microscopic artificial regimes
which electrons are forbidden to enter).~\cite{ye02} 
The microwave conductivity is enhanced with a broad peak centered 
around $\nu = 1/2$, exceeding its dc-value by as much as 
a factor of 5 for microwave frequencies up to 10 GHz. 
The enhanced conductivity is suppressed by increasing temperature 
and disappears for temperature $T > 0.5$ K.
Since the conductivity enhancement and the associated peak 
are absent for samples without antidots, 
the effect is believed to be related to the antidot edge excitations 
coupled to the microwave field. However, from
the submicron size of the antidots (with depletion regime diameter 
150 - 250 nm) one can easily estimate the usual edge magnetoplasmon modes to
have characteristic frequencies above 100 GHz, much higher than the microwave
frequencies used experimentally.
Therefore, {\em additional} lower-energy edge excitations are required 
to interpret the mysterious data.

As we discuss later in the paper, it is likely that both sets of puzzles are
related to edge reconstruction and in particular the additional edge modes
associated with it. We present further numerical evidence in support of the
generality of edge reconstruction in the fractional quantum Hall regime by
taking into account effects like layer thickness and Landau level mixing, 
which are of
quantitative importance in edge physics but not considered in our 
earlier work. We estimate the finite size effects of our
numerical studies and find that they have very little effect on our 
quantitative results of the position of reconstruction transition point.
We have also studied in this work
the electric dipole spectral
function that is directly related to the microwave conductivity
measurement and the single electron spectral function, which is what the
tunneling experiments measure. Semiquantitative comparison between our results
and experiments will be made.

The rest of the paper is organized in the following way.
We review our microscopic model in Sec.~\ref{sec:model}. 
Using heuristic electrostatic calculations, we discuss the 
origin of the reconstruction of quantum Hall edges 
and estimate the finite-size effects on microscopic model
calculations.
In Sec.~\ref{sec:diag} we present the numerical evidence 
for edge reconstruction for boundary conditions describing 
both sharp and smooth edges. 
We consider the complications due to LL mixing
and finite thickness of the quasi-2D electron layer. 
Sec. \ref{sec:finiteT} studies the effects of thermal fluctuations 
at finite temperature on edge reconstruction.
The relevance of our model and the experimentally observed 
enhanced microwave conductivity of samples with antidots
is discussed in Sec.~\ref{sec:dipole}, where we present calculations 
of electric dipole spectral functions. 
We discuss the effects of the edge reconstruction 
on the single electron spectral function 
at the edge of fractional quantum Hall liquids
in Sec.~\ref{sec:tunnel}.
We summarize our results in Sec.~\ref{sec:conclusion}.

\section{model and electrostatic considerations}
\label{sec:model}

\begin{figure}
{\centering \includegraphics{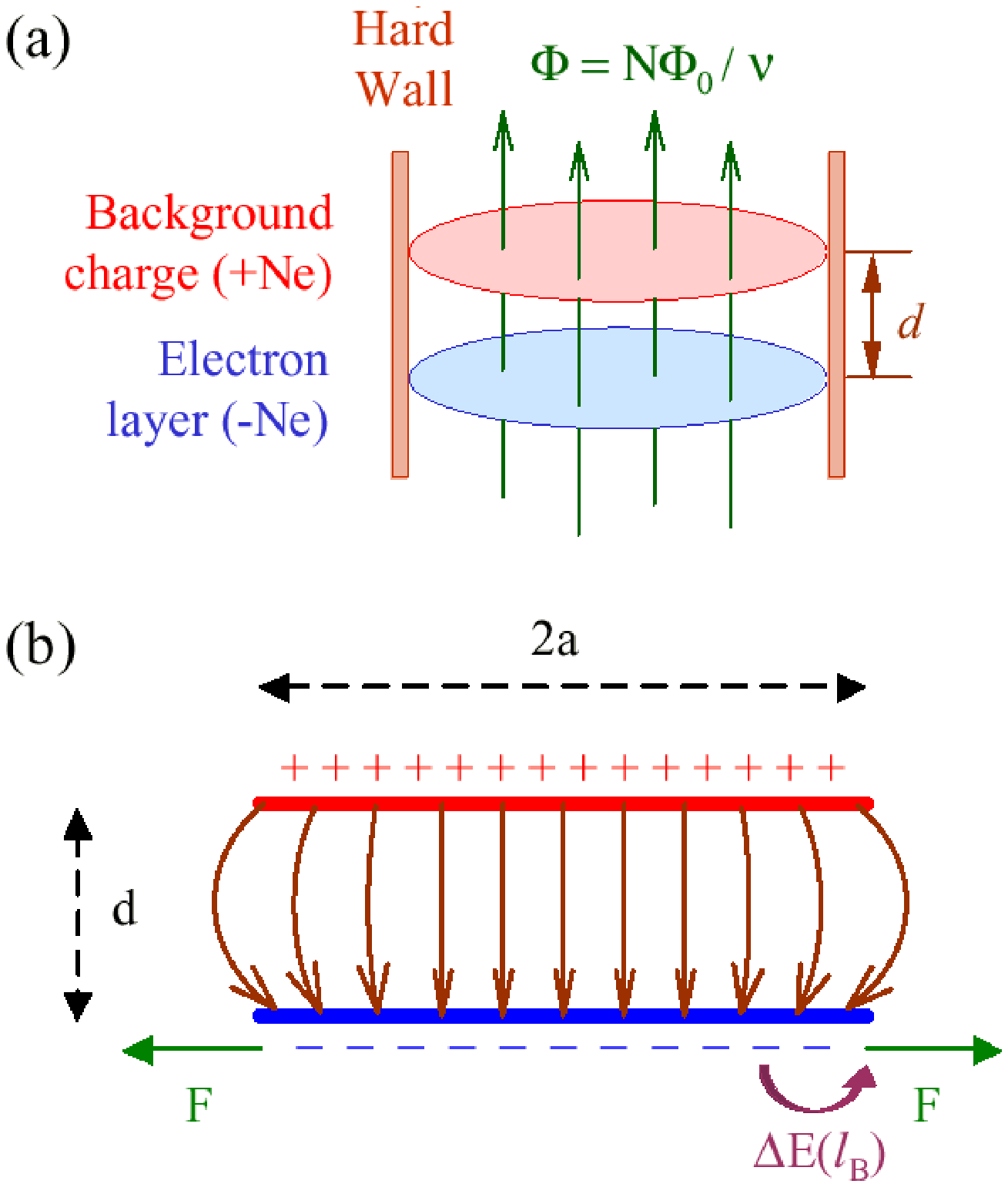} }
\caption{
\label{geometry}
(a) Sketch of the system with rotational symmetry considered here, 
which is made of an electron layer and a uniformly distributed, 
neutralizing background charge layer 
separated by a distance $d$ from each other. 
Electrons are confined by a hardwall boundary condition, so they 
cannot move beyond the edge of the background charge. 
(b) The side view of the system. 
If electrons are uniformly distributed, the electrostatic potential
is a constant in the bulk of the electron layer, but a gradient (or
fringe electric field with in-plane component) develops at the edge,
which tends to pull the electrons toward the edge.  
}
\end{figure}

We consider a two-dimensional electron gas (2DEG) with disk geometry, 
as depicted in Fig.~\ref{geometry}(a). 
To model a realistic confining potential, as in a modulation-doped
AlGaAs/GaAs heterostructure, we assume the neutralizing 
background charge is distributed uniformly on a parallel disk 
at a distance $d$ above the 2DEG. 
The radius $a$ of the positive charge background is so determined that 
the disk encloses exactly $N/\nu$ magnetic flux quanta for $N$ electrons
of the 2DEG, for any desired filling factor $\nu$. 
The bare Coulomb interaction between the background charge and the 
2DEG gives rise to the confining potential.  

Therefore we consider the following Hamiltonian, which describes
electrons confined to the lowest LL, 
using the symmetric gauge:
\begin{equation}
\label{hamiltonian}
H={1\over 2}\sum_{mnl}V_{mn}^l c_{m+l}^\dagger c_n^\dagger c_{n+l}c_m
+\sum_m U_mc_m^\dagger c_m,
\end{equation}
where $c_m^\dagger$ is the electron creation operator for the lowest LL
single electron state with angular momentum $m$. 
$V_{mn}^l$ is the matrix element of Coulomb interaction, 
\begin{equation}
\label{coulomb}
V_{mn}^l = \int d^2 r_1 \int d^2 r_2 
\phi^*_{m+l}(r_1) \phi^*_n(r_2) {e^2 \over \epsilon r_{12}} 
\phi_{n+l}(r_2) \phi_m(r_1),
\end{equation}
explicitly given by Girvin and Jach~\cite{girvin83} for symmetric gauge.\cite{note} 
$U_m$ is the matrix element of the rotationally invariant confining 
potential due to the positive background charge,
\begin{equation}
\label{confine}
U_m = {N e^2 \over \pi a^2 \epsilon} \int d^2 r_1 \int_{r_2 \le a} d^2 r_2 
{|\phi_m(r_1)|^2 \over \sqrt{r_{12}^2 + d^2}} .
\end{equation}
Here $\phi_m$ is the lowest LL wave function
\begin{equation}
\phi_m(z) = (2 \pi 2^m m!)^{-1/2} z^m e^{-|z|^2/4},
\end{equation}
where $z = x+iy$ is the complex coordinate in the plane of the 2DEG.

Before we diagonalize the Hamiltonian for microscopic systems to look 
for evidence of edge reconstruction,
we first present a heuristic argument that reveals the electrostatic origin of
edge reconstruction. 
Assuming the electrons are confined by a hard-wall boundary condition 
at $r = a$, we can view the system as a parallel disk capacitor 
in the electrostatic context, as illustrated in Fig.~\ref{geometry}(b).
If there is no edge reconstruction, to a good approximation the electron density
is uniform from the center of the disk all the way to the edge. 
Together with the uniform positive background charge,
the capacitor is uniformly charged in this case. Within the electron gas layer,
the electrostatic potential
is a constant in the bulk, but a gradient (or
fringe electric field with in-plane component) develops at the edge.
This fringe field
tends to pull the electrons toward the edge; the distance from the edge
over which the fringe field effects are significant is roughly $d$.  
Therefore, the system can gain electrostatic energy by moving electrons 
{\em outward}
near the boundary along the radial direction.
This is expected to happen when the separation between the two oppositely charged disks 
is large enough. 
When $d$ is large, the fringe field effects are strong;
thus the electrostatic energy gain from
moving some electron density outward at the edge 
overcomes the associated 
loss of exchange-correlation energy. We identify this
as the driving force of edge reconstruction.

\begin{figure}
{\centering \includegraphics{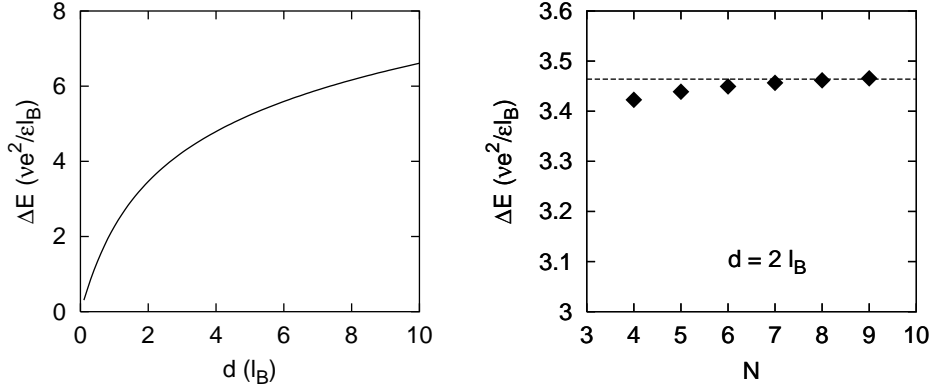} }
\caption{
\label{static}
Estimate of electrostatic energy associated with edge reconstruction and 
finite size effects.
Consider a parallel disk capacitor with uniformly distributed positive
and negative charges.
Left panel shows the potential energy gain [$\Delta E = (2d / l_B)
\tan^{-1} (l_B/d) + ln(1 + d^2/l_B^2)$] in moving an electron from one
magnetic length ($l_B$) inside the edge to the edge, for a half-infinite
capacitor system (the infinite-size limit of the disk system).
$d$ is the distance between the two charge layers.
Right panel shows $\Delta E$ in finite-size disk systems (4-9 electrons), 
with $d = 2 l_B$, at $\nu = 1/3$.
The dashed line is the infinite-size limit for $\Delta E$ at $d = 2 l_B$.
}
\end{figure}

To obtain an estimate of the energetics of this effect,
we have calculated the potential energy change of moving
one electron from one magnetic length $l_B = \sqrt{\hbar c / eB}$
(the typical length scale associated with edge reconstruction)
inside the edge to the very edge. The energy gain is
\begin{equation} 
\Delta E = (2d / l_B) \tan^{-1} (l_B / d) + \ln (1 + d^2/l_B^2), 
\end{equation}
for a half-infinite system, where $\Delta E$ is in units of
$\nu e^2/\epsilon l_B$.
This energy diverges as $2 \ln (d / l_B)$ at large $d$, as 
shown in Fig.~\ref{static}(a).
Once this energy gain exceeds the loss of exchange-correlation
energy (which must saturate in the large $d$ limit), 
edge reconstruction occurs. 
This calculation demonstrates that edge reconstruction must occur 
in infinite systems 
for
sufficiently large $d$ as long as there are states available for electron 
rearrangements. In addition, it provides an estimate of the 
energy scale associated with edge reconstruction. It also points to a 
fundamental difference between the integer and fractional bulks, in the 
limit of infinite Landau level spacing. For the former, all the possible single
electron states are occupied when an infinitely sharp edge is present; thus
no reconstruction is possible no matter how large $d$ is. On the other hand, this
is not the case for fractional bulk and reconstruction is guaranteed to occur
for sufficiently large $d$, despite the presence of a sharp edge boundary. 

In this work we perform finite size numerical studies on systems with sizes
ranging from 4 to 9 electrons.
It is important to ask whether the numerical results 
presented in this paper 
reveal the physics in the thermodynamic limit 
or merely finite-size artifacts. 
Repeating the above calculation for a finite-size disk capacitor 
can provide a measure of finite-size effects. 
The results for systems corresponding to
4 to 9 electrons are shown in Fig.~\ref{static}(b) 
for $d = 2l_B$ (which is close to critical $d_c$ above which edge reconstruction
occurs, say at $\nu=1/3$). 
First, we find that the electrostatic
energy difference between the edge and one
magnetic length inside the edge is very close to the infinite-size value
(all within 2\%). 
This suggests the finite-size effect is weak, consistent with
our finding that there is essentially no dependence of $d_c$ 
on size.~\cite{wan02}
Second, finite-size effects {\em reduce} the electrostatic energy gain 
and thus work {\em against} the edge reconstruction.
Therefore, we believe that the edge reconstruction we find is
robust and not due to finite-size artifacts. 


\section{exact diagonalization study}
\label{sec:diag}

In this section, we present the ground state properties and low energy 
excitation spectra of the 
model Hamiltonian, obtained by exact diagonalization, and look for evidence of
edge reconstruction in them. 
We discuss the appropriate choice of boundary conditions at the 
edge of the system and show that edge reconstruction is a robust property 
of the system under different boundary conditions.
In particular, we consider the effects of Landau level mixing near the edge due
to the presence of a hard-wall
potential.
We have found that the LL mixing effects, while not of qualitative
importance to the edge reconstruction physics, do affect the critical 
spacing $d_c$ quantitatively. 
Our results are also robust in the presence of finite thickness of 
an electron layer, whose effects turn out to be negligible as long as 
we treat the electron layer as a sheet of charge at its maximum
density. 

\subsection{Sharp edge}

\begin{figure}
{\centering \includegraphics{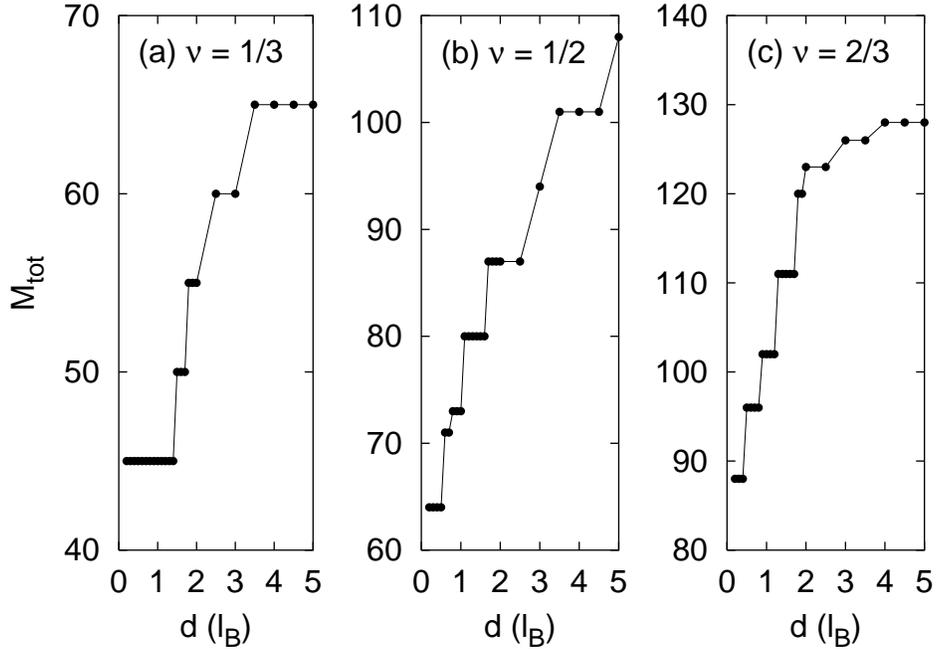} }
\caption{
\label{gsnumber}
Dependence of $M_{tot}$ on $d$ for 6, 9, and 12 electrons in 18
orbitals, which correspond to $\nu =$ 1/3, 1/2, and 2/3,
respectively. 
}
\end{figure}

In the previous study,~\cite{wan02} 
we restricted electrons to $N / \nu$ orbitals of the lowest LL ({\em i.e.},
no Landau level mixing), 
from $m = 0$ to $m_{max} = N / \nu - 1$.
Such a constraint was introduced to describe, for filling factor $\nu$,
the presence of a sharp cleaved edge~\cite{pfeiffer90} 
beyond which electrons cannot move. 
In the exact diagonalization study, we found, for $\nu = 1/3$, 
the following evidence that supports the edge reconstruction scenario.
(1) The total angular momentum of the global ground state $M_{tot}$ 
becomes greater than that of the corresponding Laughlin state
when the separation $d$ between electron and background charge layers 
exceed critical $d_c = 1.5 \pm 0.1 l_B$; the increase of $M_{tot}$ is due to 
the outward motion of electrons near the edge triggered by the reconstruction,
as discussed in the previous section.
The critical value $d_c$ is essentially the same
for $N = $ 4-9 electrons, with very little
size dependence. 
(2) The electron density profile for the global ground state shows
significant oscillation near the edge 
for $d > d_c$.
(3) Counterpropagating low-lying excitations can be identified in 
the low-energy excitation spectrum for $d > d_c$. 
We also found that property (2) is present for filling factors 
$1/3 \le \nu \le 2/3$, suggesting edge reconstruction is generic for 
fractional bulk fillings. Properties (1) and (3), on the other hand,
are unique to $1/3$:
(1) because of the existence of the Laughlin state as the reference state 
and (3) because the single chiral boson mode of the unreconstructed edge
does not, in general, extend to other filling factors. So, the mere existence
of counter-propagating modes 
can not be used for general
filling factors to look for edge reconstruction. 
Nevertheless, the general trend remains that $M_{tot}$ increases with
$d$ for both incompressible and compressible filling fractions,
indicating that fractional QH edges tend to reconstruct 
for large $d$. 
Figure~\ref{gsnumber} shows the dependence of $M_{tot}$ on $d$ for 6, 9,
and 12 electrons in 18 orbitals, or $\nu =$ 1/3, 1/2, and 2/3,
respectively. 
In all three cases, $M_{tot}$ starts to jump at $d \sim 1 l_B$. 

These numerical findings are consistent with the heuristic arguments
that we gave in the previous section. 
One key quantity in these calculations is the critical $d$, which has 
always been found to be of the order of $l_B$ 
(even for the more complicated cases we discuss later in this section).
We believe that this is not a coincidence. 
Within the lowest LL, $l_B$ is a
fundamental length scale which characterizes the size of single
electron wave functions. 
It is thus also the range of effective attraction between electrons due to 
exchange-correlation effects and therefore the length scale associated with
edge reconstruction.
On the other hand, the separation between the
electron and background charge layers, $d$,
is the range of the electrostatic fringe field near the
edge.
Therefore, the electrostatic energy gain generally becomes comparable to
the loss of
electron exchange-correlation energy when $d$ 
is of the order of $l_B$; we thus expect $d_c\sim l_B$. 

\begin{figure}
{\centering \includegraphics{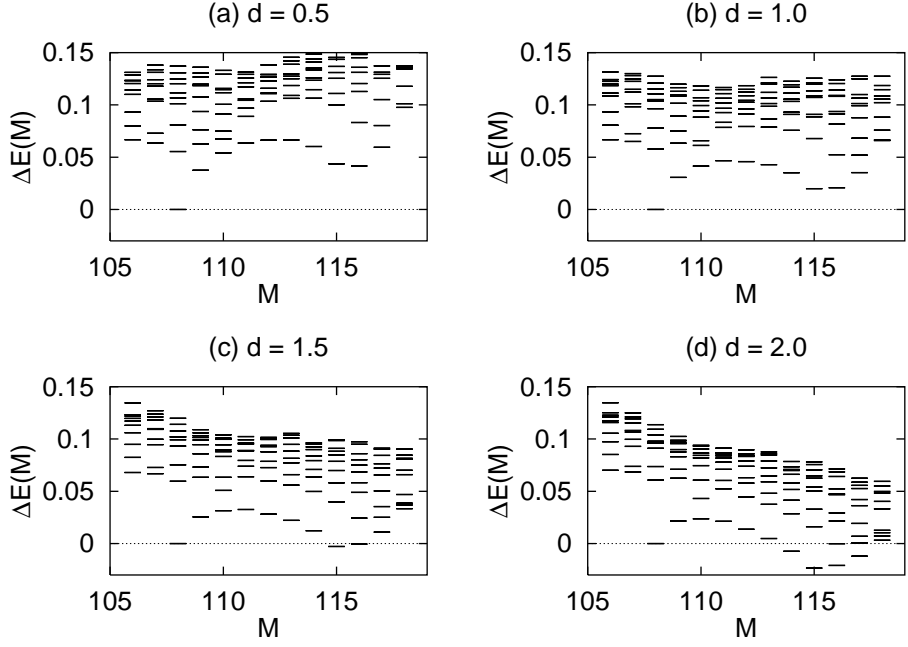} }
\caption{
\label{spectra0}
Low energy spectra for $N = 9$ electrons in 27 orbitals
for (a) $d = 0.5$, (b) $d = 1.0$, (c) $d = 1.5$,
and (d) $d = 2.0$, in units of $l_B$.
Excitation energies ($\Delta E$) are measured, 
in units of $e^2 / \epsilon l_B$,
from the ground state in the $M_{tot} = 108$ 
(that of the corresponding Laughlin state) subspace. 
The edge reconstruction transition occurs around $d = 1.5$, as 
$\Delta E$ becomes negative for $M_{tot} = 115$. 
}
\end{figure}

Additional insight into the physics of edge reconstruction may be gained by
studying how the low-energy excitation spectrum of the system evolves as $d$
increases. According to the CLL theory, for simple bulk fillings like 1/3, 
there exists a single branch of 
chiral bosonic edge modes with linear dispersion
in the long wave-length limit in the absence of edge reconstruction; 
these modes describe the propagation of 
the deformation of the periphery of a QH system. 
This branch of chiral bosonic edge modes
is clearly visible in Figure~\ref{spectra0}, where we show
the low-lying spectrum for an $N = 9$ system
at $\nu = 1/3$ (or, $m_{max} = 26$), for both $d < d_c$ and $d > d_c$.
The chirality of the mode is reflected in the fact that low-lying excitations 
exist only in subspaces $M > M_0$, where $M_0$ is the quantum number of the 
ground state (which is the same as that of the Laughlin state) for $d < d_c$.
What is worth emphasizing here are two features not described by the CLL 
theory.
First, the low-lying edge modes, which are clearly separated from the bulk
excitations with a gap, exist for relatively large $\delta M = M-M_0$ 
(corresponding
to large momentum or short wave-length in translationally invariant systems). 
From the system size we can estimate the momentum
$k = \Delta M/2\pi R \approx 0.2 l_B^{-1}$ for which 
the mode remains well-defined.
Second, the dispersion relation shows significant deviation from linearity and
even becomes non-monotonous for larger momenta, with a well-defined local 
minimum near $k\approx 0.15 l_B^{-1}$. This non-linearity grows with
increasing $d$. At $d=d_c$, the mode energy crosses zero at this local
minimum. Thus the edge reconstruction transition can also be understood as 
an instability, driven by $d$, of the (unreconstructed) chiral boson mode at finite wave 
vector (of order $l_B^{-1}$). 
This clearly
indicates that edge reconstruction is driven by short-distance physics at the
scale of $l_B$, in agreement with previous analysis. Based on this insight,
a unified field theoretical description for the edges with and without 
reconstruction, and the transition between the two is advanced by one of 
us.\cite{yang03}

As discussed earlier, a
cutoff at $m_{max}$ for single electron states is introduced to model a sharp
cleaved edge. 
Such a sharp edge is not present for all samples.
To describe a smooth edge (where confinement is provided only
by the positive background), we can move the cut-off to 
higher angular momentum.
In fact, even a sharp cleaved edge is more complicated than 
imposing a simple cut-off at $m_{max}$.
More appropriately, electrons are confined in a disk 
of radius $a$, with a very high step-potential at the edge 
$r = a$. 
The confinement has two related effects.
First, the step-potential mixes in high Landau level components for single
electron wave functions near the edge and raises the energies for 
single-particle states, even with angular momentum
$m < m_{max}$ (this is in addition to the effect of the back ground charge).
Second, the change of single-particle wave function,
caused by the in-mixing of LLs, changes essentially 
all the Coulomb interaction matrix elements in 
Eq.~(\ref{hamiltonian}).
These effects are to be considered in the following 
subsections.

\subsection{Smooth edge}

\begin{figure}
{\centering \includegraphics{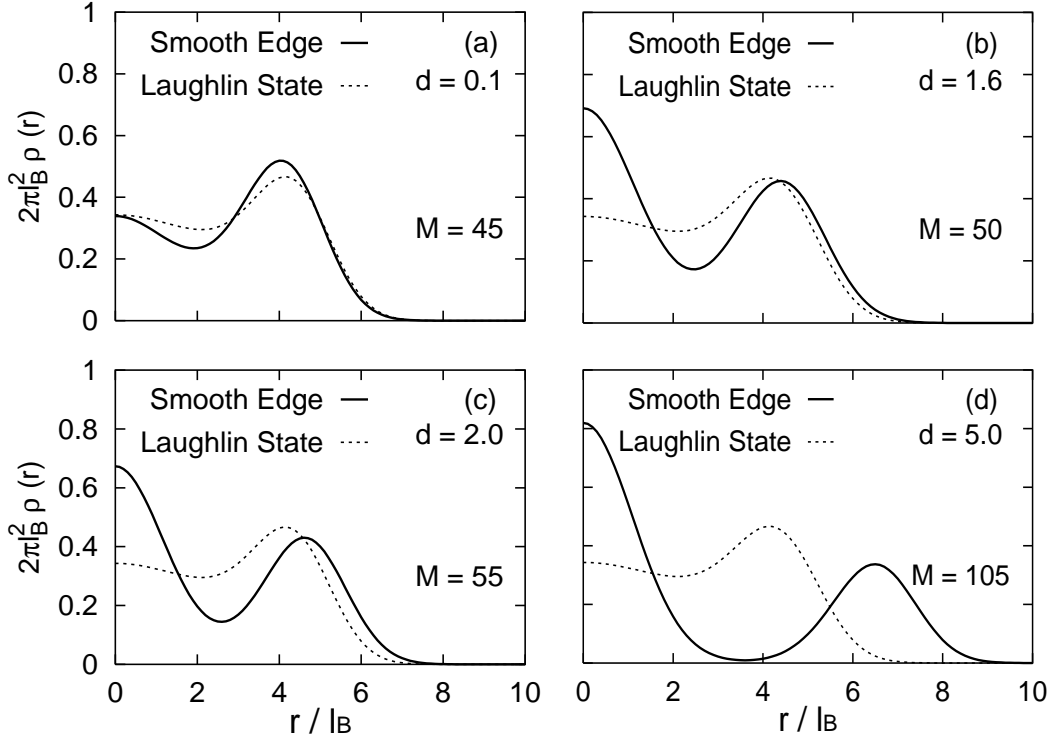} }
\caption{
\label{smoothEdge}
The electron density $\rho(r)$ of the global ground state 
for 6 electrons in 30 orbitals compared with that of the Laughlin
state (dotted lines) for (a) $d = 0.1$, (b) $d = 1.6$, (c) $d = 2.0$, 
and (d) $d = 5.0$, in units of $l_B$.  
The radius of the disk with uniform background charge corresponds 
to $\nu = 1/3$, so electrons are allowed to move $\sim 2 l_B$ beyond 
the background charge.  
}
\end{figure}

As mentioned in the previous subsection, 
while a sharp cutoff (at $m_{max}$)
where the background charge ends describes samples with
cleaved edges, it is not present for other samples. 
Not surprisingly, we find that removing this
sharp cutoff can only further favor edge reconstruction. 
For the 6-electron system considered before,~\cite{wan02} 
we have increased the number of orbitals to 30 ($m_{max} = 29$), 
while keeping the background charge unchanged 
so that the sharp cutoff is moved
about $2 l_B$ away from the boundary of the background charge.\cite{cutoffnote} 
Figures~\ref{smoothEdge}(a)-(d) show electron densities of the 
global ground states for $d = 0.1$, 1.6, 2.0, and 5.0 in the presence
of such a smoother edge.  
Compared with Fig.~1(b)-(e) in Ref.~\onlinecite{wan02}, 
we find that the change of
the sharp cutoff has essentially no effect on the electron density for $d
\le 2.0$. 
In particular, the critical distance $d_c = 1.6$ remains
very close to that of the sharp edge case.  
In these cases, we expect that the electronic states are almost entirely
determined by the competition of the electron-electron interaction and
the confining potential arising from the background charge. 
However, with the cutoff moved farther out, the edge piece can shift
further away from the disk center for larger $d$ so that the total
momentum $M_{tot}$ of the lowest ground state increases, 
e.g. from $M_{tot} = 65$ to 105 for $d = 5.0$. 
Therefore, the inclusion of extra single electron orbitals
has very little effect on the edge reconstruction transition; 
the effects are important only
for systems well in the reconstructed phase with 
$d \gg d_c$. 
For simplicity, in the following discussion we do not include
these additional orbitals beyond the boundary 
of the background charge distribution unless otherwise specified.

\subsection{Layer thickness}

The electrons confined at the interface of the modulation-doped 
Al$_x$Ga$_{1-x}$As/GaAs heterostructures, as used in experiments, 
are not ideally two-dimensional (2D) although their motions
perpendicular to the interface are essentially 
frozen in their ground state due to sharp interface potential. 
In principle, one needs to use the self-consistently calculated wave
function appropriate for the heterostructures to study the softening 
of electron-electron interaction due to finite electron layer 
thickness. 
However, Stern and Das Sarma~\cite{stern84}
showed that the Fang-Howard variational wave function,~\cite{fang66} 
\begin{equation}
\label{fang-howard}
Z_0(z) = 2 (2b)^{-3/2} z e^{-z/2b},
\end{equation}
is a very good approximation to the numerical self-consistent 
ground state. 
The Fang-Howard wave function $Z_0(z)$ peaks at $z_0 = 2b$. 
The parameter $b$ gives the scale of the layer thickness, which is 
typically $\sim 50$ \AA. 
The finite layer thickness weakens the electron-electron 
interaction, as well as the background charge 
confining potential.
For instance, the effective electron-electron interaction 
in the quasi-2D system is approximated by 
\begin{equation}
V(|\vec{r}_1 - \vec{r}_2|) = {e^2 \over \epsilon} 
\int\!\!\!\int dz_1 dz_2 { |Z_0 (z_1)|^2 |Z_0 (z_2)|^2 \over
[r^2 + (z_1 - z_2)^2]^{1/2} },
\end{equation}
where $r$ is the in-plane distance between two electrons. 

\begin{figure}
{\centering \includegraphics{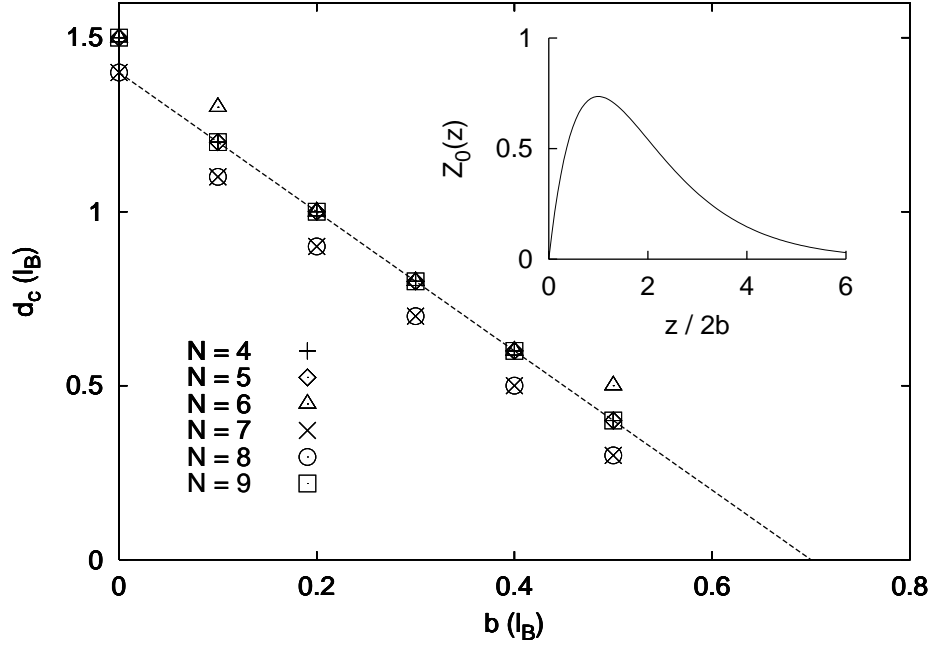} }
\caption{
\label{layerthickness}
Critical $d_c$ (in units of $l_B$) for edge reconstruction, 
at which the ground state momentum $M_{tot}$ becomes greater 
than that for the corresponding Laughlin state,
for $N$ = 4-9 electrons at $\nu = 1/3$ with finite layer thickness.
$d_c$ is measured as the distance from the positive background charge layer
to the (GaAs/AlGaAs) interface where potential is discontinuous.
The finite thickness of the 2D electron gas in the perpendicular direction is
described by the Fang-Howard variational wave function
$Z_0(z) = 2 (2b)^{-3/2} z e^{-z/2b}$ (see inset).
$d_c$ can be roughly fit to $d_c = 1.4 - 2b$, where $1.4 \pm 0.1$ can be
regarded as the critical $d$ for zero layer thickness, and $2b$ the
distance from the peak of the variational wave function to the interface
($z = 0$). 
}
\end{figure}

For typical experimental parameters, the electron wave functions 
have a finite thickness $2b \sim$ 100 \AA, equal to roughly 
one magnetic length.  
For an AlGaAs/GaAs heterojunction, we choose the $z$-direction 
pointing from the AlGaAs to the GaAs 
with the AlGaAs/GaAs interface placed right at the $z = 0$ plane.
Since the AlGaAs introduces a potential barrier, the quasi-2D
electron density spreads essentially in the GaAs.
The background charge layer, introduced by $\delta$-doping, 
is thus located at the $z = -d$ plane (on the AlGaAs side).
Experimentally, $d \sim$ 800 \AA\ or above.
One central question, therefore, is whether the edge reconstruction 
persists in the presence of the finite layer thickness;
in particular, whether $d_c$ for the edge reconstruction
remains smaller than the typical $d$ used in experiments. 
To answer this, we repeat exact diagonalization calculations 
for several values of $b$, 
searching for the critical $d$ above which the global 
ground state has a larger total angular momentum than that of 
the corresponding Laughlin state. 
In this calculation, we use a sharp cutoff ($m_{max} = N / \nu - 1$)
in the angular momentum space and do not consider the complication 
of LL mixing. 
Figure~\ref{layerthickness} summarizes the results for $N =$ 
4-9 electrons for $\nu = 1/3$. 
Again, $d_c$ is almost size-independent for $N =$ 4-9 electrons, 
confirming that finite-size effects are weak in this calculation.
Overall, $d_c$ decreases as $b$ increases, and can be 
roughly fit by 
\begin{equation}
d_c = d_c^0 - 2b,
\end{equation}
where $d_c^0 = 1.4 \pm 0.1$.
Note $d_c$ decreases slightly faster than $2b$ for small $b$;
thus $d_c^0$ is smaller than $d_c = 1.5 \pm 0.1$, which is found 
for systems with zero thickness. 
Since $2b$ is the distance between the AlGaAs/GaAs interface 
and the peak of the Fang-Howard wave function, an alternative 
interpretation of the results is that $d_c$ remains roughly
as a constant, if we measure $d$ from the background charge layer
to the plane of maximum electron density, instead of the interface. 
Thus the overall effect of finite layer thickness favors edge reconstruction,
and reduces the critical dopant layer distance $d_c$ slightly.

\subsection{Hard-wall confinement and Landau level mixing near the edge}

So far, we have been working in the limit where the kinetic energy 
is completely quenched, so electrons are entirely in the lowest LL. 
In this limit, the Coulomb energy $e^2 / \epsilon l_B$ is the 
only energy scale in the system. 
However, for typical $n$-type GaAs heterostructures in experiments, 
the Coulomb energy is comparable to the cyclotron energy 
$\hbar \omega_c$ separating the LLs. 
For the fractional QH effect, the LL mixing effects 
have been considered in numerical as well as analytical 
studies.~\cite{yoshioka86,melik95,melik97,murthy02}

\begin{figure}
{\centering \includegraphics{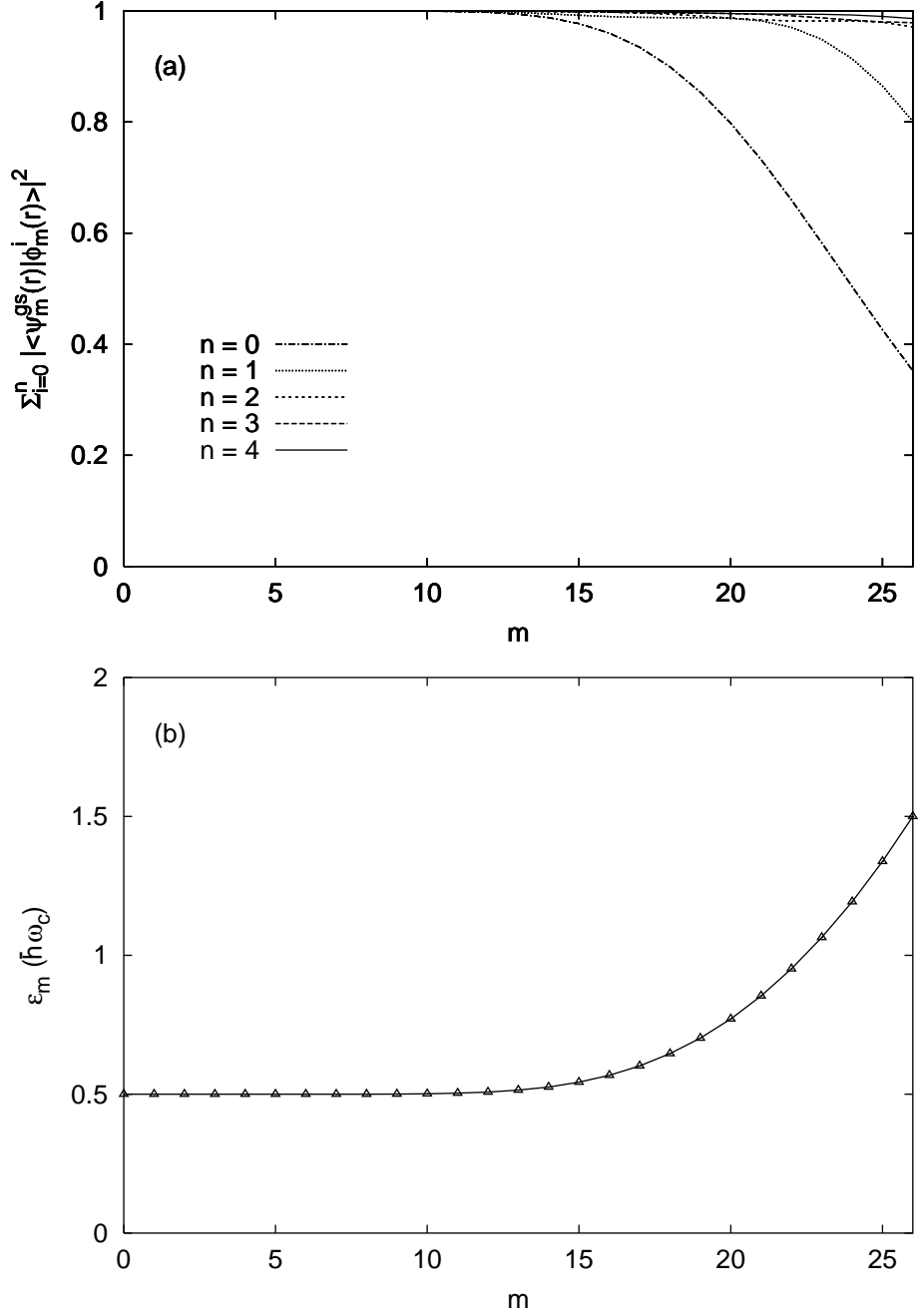} }
\caption{
\label{llmixing}
(a) Cumulative overlaps, 
$\sum_{i=0}^n \left | \langle \psi_m^{gs}(r) | \phi_m^i(r) \rangle \right
|^2$, of the ground state wave function [$\psi_m^{gs}(r)$] for each
angular momentum  $m$, in the presence of a hard-wall boundary
condition, with LL wave function, $\phi_m^i(r)$, 
for the lowest five LLs ($i$ = 0-4). 
The sum of the overlaps is more than 99\% for
each $m$. 
(b) The single-particle energy of the ground state $\epsilon_m$
increases from $\hbar \omega_c / 2$ in the bulk to 
$3 \hbar \omega_c / 2$ at the edge. 
}
\end{figure}

The LL mixing effects are important for edge physics, 
since the single particle energy for electrons confined in a disk 
increases monotonically from bulk to edge and 
eventually crosses higher LL energies due to confinement, 
as first discussed by Halperin~\cite{halperin82} in the 
QH context. 
Therefore, LL mixing, in particular, resulting from 
the cleaved sample edge may as well alter the edge physics. 
To include this effect, we solved the Schr\"odinger equation 
for non-interacting electrons confined in a two-dimensional disk 
(we neglect the finite thickness of the electron layer) 
by a hard-wall boundary condition:
\begin{equation}
\psi(r) = 0, \ {\rm for} \ r > a.
\end{equation}
The ground state wave function $\psi^{gs}_m (r)$ in each angular momentum 
($m$) subspace 
now becomes a mixture of states in all
LLs with the same $m$ quantum number. 
As a result, the energy of the ground states increases 
from $\hbar \omega_c / 2$ (lowest LL value) for $m = 0$ to  
approach $3 \hbar \omega_c / 2$ (first LL value) 
for $m_{max} = 3N -1$, as depicted in Fig.~\ref{llmixing}(b) (here we do not 
include contributions from the confining potential of the background charge). 
Figure~\ref{llmixing}(a) shows the cumulative overlaps, 
$\sum_{i = 0}^n \left | \langle \psi^{gs}_m (r)| \phi^i_m (r)\rangle \right
|^2$, of the corresponding ground state wave functions 
[$\psi^{gs}_m (r)$] for each angular momentum  $m$, 
with the LL wave function [$\phi^i_m(r)$] for the lowest five LLs 
($i$ = 0-4). 
The total overlap of the five lowest LLs is more than 99\% 
for each $m$. 

Following the usual procedure of projecting the Hamiltonian of 
the system onto the ground state manifold, we obtain an effective 
Hamiltonian in the same form as Eq.~\ref{hamiltonian}, with 
$c_m^\dagger$ creating an electron in the LL-mixed ground state
with angular momentum $m$. 
The single particle wave functions $\phi_m$ in Eq.~(\ref{coulomb}) and
(\ref{confine}) are replaced by $\psi^{gs}_m$.
We absorb the $m$-dependent single-particle energy $\epsilon_m$ into 
the confining energy term $U_m$.  
We restrict the Hilbert space to $0 \le m < 3N$ for $\nu = 1/3$. 
Note that we have, in addition to $e^2 / \epsilon l_B$, a second 
energy scale $\hbar \omega_c$. 
The LL mixing effect is thus characterized by the dimensionless 
parameter $\lambda = (e^2 / \epsilon l_B) / \hbar \omega_c$. 
We are, in particular, interested in $\lambda \sim 1$, which is close 
to the real experimental conditions.
The LL mixing raises $\epsilon_m$ by roughly $\hbar \omega_c$ 
for edge states, thereby  making electrons occupying these states
energetically unfavorable. 
However, the squeezing of the wave functions near the edge due to the Landau 
level mixing effect 
lowers the confining potential from the background charge 
and also reduces the range of effective attraction due to exchange-correlation 
effects. These effects favor edge reconstruction.

\begin{figure}
{\centering \includegraphics{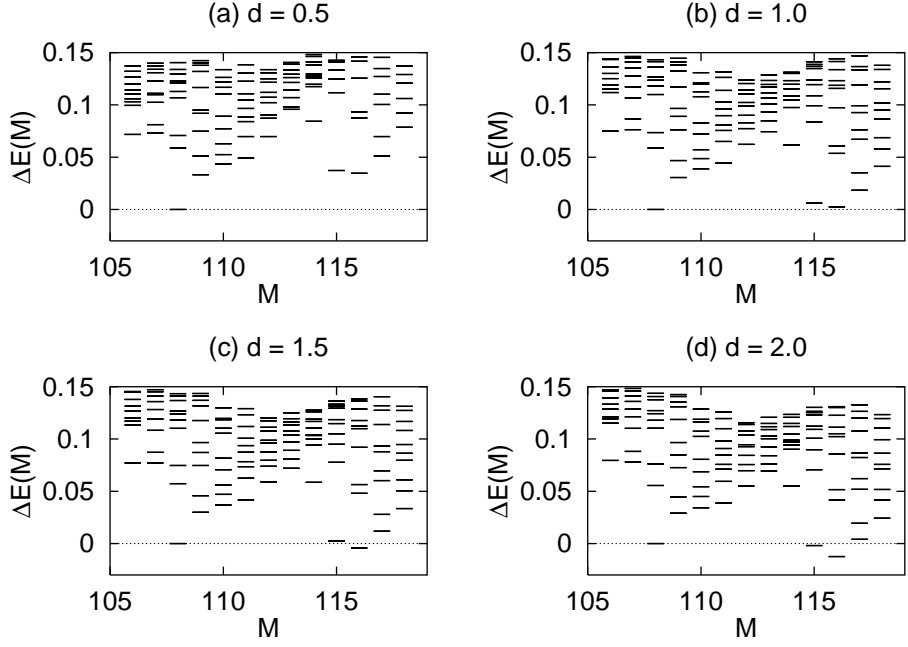} }
\caption{
\label{spectra}
The low energy spectra for $N = 9$ electrons in 27 orbitals with
hard-wall boundary conditions for 
(a) $d = 0.5 l_B$, with no edge reconstruction,
(b) $d = 1.0 l_B$, close to but before edge reconstruction,
(c) $d = 1.5 l_B$, and 
(d) $d = 2.0 l_B$, both after edge reconstruction.
Excitation energies ($\Delta E$) are measured, 
in units of $e^2 / \epsilon l_B$,
We choose the dimensionless parameter 
$\lambda = (e^2 / \epsilon l_B) / \hbar \omega_c = 2.0$ here.
After edge reconstruction, the total ground state momentum 
becomes $M_{tot} = 116$, increasing from $M_{tot} = 108$ 
of the corresponding Laughlin state.
}
\end{figure}

Figure~\ref{spectra} shows the low-energy spectra for $N = 9$
electrons in 27 orbitals with hard-wall boundary conditions
for various $d$ and $\lambda = 2.0$.
After edge reconstruction, the total ground state momentum 
becomes $M_{tot} = 116$, increasing from $M_{tot} = 108$ 
of the corresponding Laughlin state.
Counterpropagating low-energy modes can be observed near 
$M_{tot} = 116$ for $d = 1.5 l_B$. 
Qualitatively, these results agree very well with the scenario
using sharp cutoff in angular momentum space 
(Fig.~\ref{spectra0}).
However, we also observe fluctuations in the critical $d_c$ 
for systems with different sizes. 
For $\lambda = 1.0$, we list $d_c$ in Table~\ref{dc}, 
which varies from 0.4 $l_B$ to 1.6 $l_B$. 
Although the data seems to stabilize at $d_c \approx 1.0$, 
we cannot draw definitive quantitative conclusions without data from
larger systems in contrast to the case of no LL
mixing. We believe the increased finite size effect here is due to 
the fact that LL mixing affects states in a relatively wide region near the
edge; this effect was not present in our earlier study.
We also point out that here we define $d_c$ as the distance at which 
$M_{tot}$ exceeds the corresponding value in the 
$d \rightarrow 0$ limit and the results presented are based on this working
definition. 
With strong LL mixing effects, the ground state for 
$d < d_c$ may not have an $M_{tot}$ consistent with the value of 
the corresponding Laughlin state. 
We believe that such a different $M_{tot}$ in the limit of 
$d \rightarrow 0$ suggests the Laughlin state may not be a good
approximation to the ground state. 
In other words, we probably cannot identify the finite systems 
as having filling fraction 1/3 unambiguously.

\begin{table}
\begin{center}
\begin{tabular}{c@{\hspace{0.2in}}c@{\hspace{0.2in}}
c@{\hspace{0.2in}}c@{\hspace{0.2in}}c@{\hspace{0.2in}}c}
\hline \hline
 $N$  & 5 & 6 & 7 & 8 & 9 \\ \hline 
$d_c$ & 0.4 & 1.6 & 0.6 & 0.9 & 1.0 \\ 
\hline\hline
\end{tabular}
\end{center}
\caption{
\label{dc}
$d_c$ for edge reconstruction for 
$\lambda = (e^2 / \epsilon l_B) / \hbar \omega_c = 1.0$.
We define $d_c$ as the distance at which 
$M_{tot}$ of the global ground state 
exceeds the corresponding value in the $d \rightarrow 0$ limit.}
\end{table}

We would like to emphasize that all of our numerical results, as well as
the heuristic arguments based on electrostatic considerations, suggest
that $d_c\sim l_B$. In real samples, on the other hand, $d \gtrsim
10l_B$.  We thus believe that it is safe to conclude that the edges of
all samples studied so far are reconstructed when the bulk filling is
fractional, regardless of whether the edges are cleaved or not.

\section{Effects of finite temperature on edge reconstruction}
\label{sec:finiteT}

\begin{figure}
{\centering \includegraphics{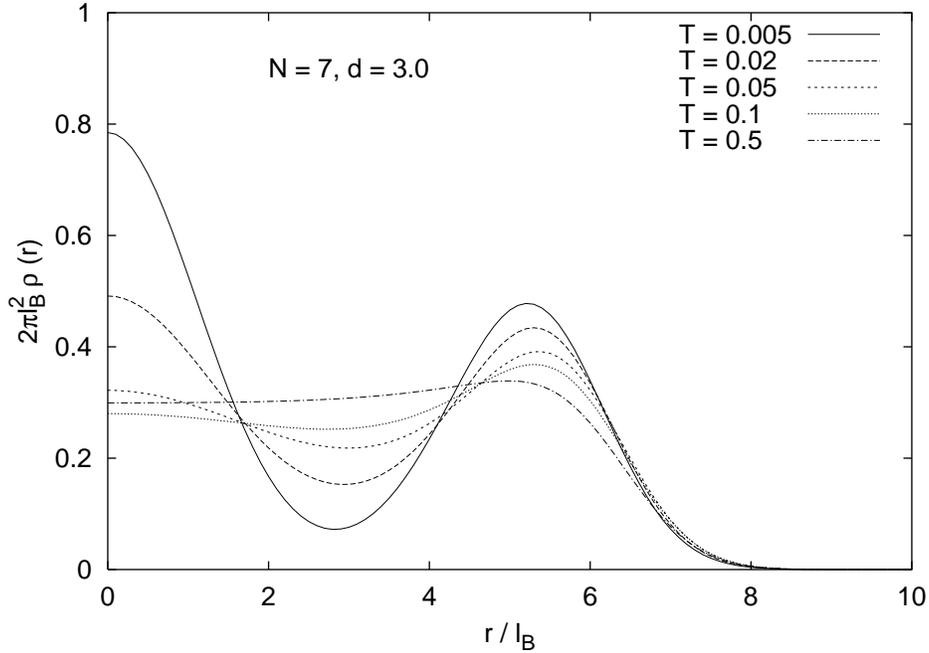} }
\caption{
\label{finiteT}
Finite-temperature (in unit of $e^2/\epsilon l_B$)
electron density profiles $\rho (r)$ for $N = 7$ electrons
at $\nu = 1/3$.
The distance between the two charge layers is fixed at $d = 3l_B$.
At and above $T = 0.05 e^2/\epsilon l_B$, $\rho(r)$ 
become similar to that of the corresponding Laughlin state,
fluctuating slightly around $1/3$.
This is very different from $\rho (r)$ at $T = 0.005 e^2/\epsilon l_B$, 
where strong density oscillation due to
edge reconstruction can be seen.
}
\end{figure}

Thus far our studies have been focusing on ground state (or zero temperature)
properties of the system, especially those related to edge reconstruction. 
It is of interest to investigate how finite temperature, and the thermal 
fluctuations associated with it, affect these properties. 
In our previous study,\cite{wan02} we used a finite-temperature Hartree-Fock
approximation to show that the reconstruction of a $\nu = 1$
QH edge is suppressed above a certain temperature $T^*$. This is quite reasonable
since thermal fluctuations tend to suppress edge density oscillations associated
with reconstruction.  
In that study the temperature $T^* \approx 0.05 e^2 / \epsilon l_B$
(or $T^* \approx 6$ K) for typical experimental parameters.
It is expected, however, that the Hartree-Fock calculation tends to
overestimate the temperature scale, due to effects of finite size,
layer thickness, LL mixing, disorder, etc. Here we study the effects of 
finite $T$ for fractional bulk filling $\nu=1/3$. In this case a Hartree-Fock 
calculation is no longer possible; we thus use exact diagonalization to 
generate all the excited states with significant thermal weight and perform
thermal averaging. 
Figure~\ref{finiteT} shows the evolution of the density profile
for 7 electrons at $\nu=1/3$ at several temperatures.
We find that, similar to what we found earlier for $\nu=1$,
the edge density oscillation associated with 
edge reconstruction is washed out above a certain characteristic temperature,
$T^* \approx 0.05 e^2 / \epsilon l_B$, at which the density profile no longer
has strong oscillations and
becomes quite similar to that of the corresponding Laughlin state.
This is about the same temperature scale (but for much smaller systems)
that we got for the integer edge from Hartree-Fock; we expect
the temperature scale will become much lower
for a comparable size system.

\section{microwave conductivity and its temperature dependence}
\label{sec:dipole}

The microwave conductivity related to microwave absorption 
of our microscopic system can be calculated through 
the electric dipole spectral function as follows:
\begin{equation}
\sigma (\omega) \propto \sum_{i \rightarrow f} 
| \langle \psi_f | \hat{e} \cdot \vec{r} | \psi_i \rangle |^2
\delta \left ( E_f - E_i - \hbar \omega \right ) 
f(E_i) \left (E_f - E_i \right ) ,
\label{cond}
\end{equation}
where $f(E)=e^{-E/k_BT}/Z$ is the thermal weight of a state with energy $E$. 
$E_{i,f}$ and $\psi_{i,f}$ are the energies and wave functions for 
initial and final states, respectively, and the summations are over 
all eigenstates. 
$\hat{e}$ is the unit vector of an electric field 
and $\vec{r}$ the position vector for electrons. 

\begin{figure}
{\centering \includegraphics{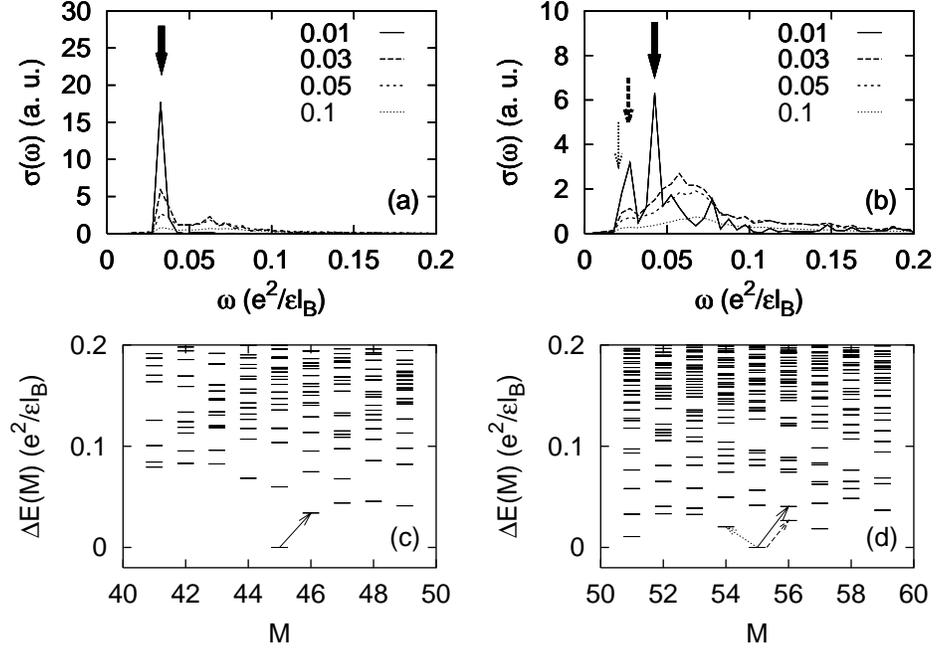} }
\caption{
\label{dipole}
Microwave conductivity $\sigma(\omega)$ calculated by electric dipole 
spectral function for $N = 6$ electrons at $\nu = 1/3$ for
(a) $d = 1.0 l_B$ (with no edge reconstruction) and
(b) $d = 2.0 l_B$ (after edge reconstruction) at various temperatures.
The corresponding low-energy excitation spectra are plotted for 
(c) $d = 1.0 l_B$ and (d) $d = 2.0 l_B$.
Before edge reconstruction, the absorption is dominated by long wave-length
chiral edge mode [solid arrows in (a) and (c)] 
even above $T = 0.05 e^2/\epsilon l$.
After reconstruction, extra modes 
[indicated by three arrows in (b) and (d)] 
due to reconstructed edge can be identified
as contributing to the peaks in $\sigma(\omega)$ for $\omega < 0.05$,
in units of $e^2/\epsilon l$.
At $T > 0.03 e^2/\epsilon l$, these modes become less significant, compared to
bulk excitation.
}
\end{figure}

The microwave conductivity, as determined from Eq.~(\ref{cond})
for a system of 6 electrons 
for $\nu = 1/3$ at different temperatures, both before ($d = 1.0 l_B$) and 
after edge reconstruction ($d = 2.0 l_B$), 
are shown in Figs.~\ref{dipole}(a) and (b). 
The data is coarse-grain averaged in frequency space for clarity of viewing. 
For $d = 1.0 l_B$, the spectral function shows a dominant peak at 
$\omega \approx 0.04$, in units of $e^2 / \hbar \epsilon l_B$. 
This peak corresponds to the single chiral edge mode 
for the principal filling factor $\nu = 1/3$, 
marked by a solid arrow in the corresponding excitation spectrum, 
Fig.~\ref{dipole}(c). 
At higher temperature, the peak becomes less prominent, 
due to the reduced statistical weight of the ground state and other low-lying 
states; however the position of the peak does not shift with changing $T$, and 
the peak can still be identified as the dominant one in the whole spectrum
for temperature up to $T = 0.05$, in units of $e^2 / \epsilon l_B$. 
For $d = 2.0 l_B$, where edge reconstruction has occurred, the 
frequency dependence of the conductivity becomes qualitatively different. 
Two distinct peaks can be resolved for $\omega < 0.05$
at $T = 0.01$, contributed by three dominant modes, 
with two in the lower peak (marked by a dotted arrow and a dashed arrow). 
The additional modes are due to edge reconstruction, 
which creates two counter-propagating
edge modes. 
What is more interesting is the manner in which the conductivity evolves as $T$
increases; here we find,
at higher temperatures, the peak of the spectral function 
{\em shifts} to higher frequencies, and the low frequency response due to the
additional modes get {\em suppressed}. 
The low-temperature peaks can barely be resolved at $T = 0.05$. 
Such behavior is in good qualitative agreement with the microwave experiments
in samples with an array of antidots (reported in Ref. \onlinecite{ye02}). 
These authors find enhanced low frequency (much lower than the edge magneto
plasmon frequency) conductivity at low $T$, while such
enhancement gets suppressed at higher $T$. One should note, however, that the
system size of our study (as, say, parameterized by the circumference of the
edge) is much smaller than those of the real samples; thus, not
surprisingly, 
the energy and temperature scales obtained in our work is considerably 
larger than those of the experimental data; this is purely a finite size
effect. On the other hand the temperature scale obtained here is consistent with
what we obtained in the previous section through the temperature dependence of
edge density oscillation.

\begin{figure}
{\centering \includegraphics{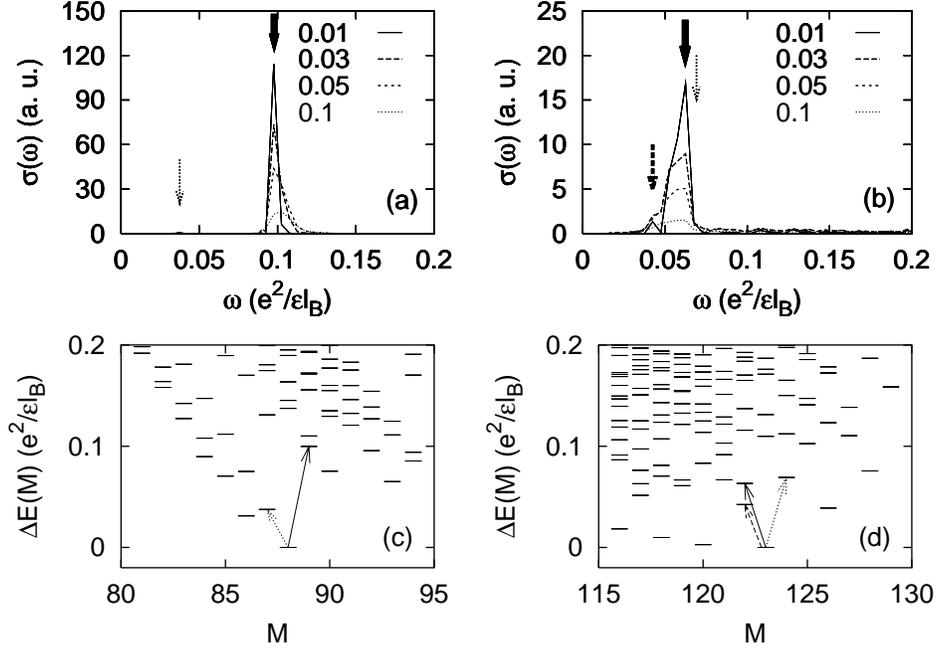} }
\caption{
\label{dipole2}
Microwave conductivity $\sigma(\omega)$ calculated by electric dipole 
spectral function for $N = 12$ electrons at $\nu = 2/3$ for
(a) $d = 0.1 l_B$ and (b) $d = 2.0 l_B$ at various temperatures.
The corresponding low-energy excitation spectra are plotted for 
(c) $d = 0.1 l_B$ and (d) $d = 2.0 l_B$.
For $d = 0.1 l_B$, $\sigma(\omega)$ is dominated by the lowest 
$\Delta M = +1$ edge mode [solid arrows in (a) and (c)],
while the lowest $\Delta M = -1$ edge mode [dotted arrows in (a) and (c)]
is significantly weaker. 
For $d = 2.0 l_B$, $\sigma(\omega)$ is dominated by the second lowest 
$\Delta M = -1$ edge mode [solid arrows in (b) and (d)]. 
Both the lowest $\Delta M = \pm 1$ modes are much weaker, 
as indicated by dashed and dotted arrows in (b) and (d).
}
\end{figure}

The microwave experiment on samples with antidots~\cite{ye02} 
suggests that the observed enhanced 
conductivity is a generic feature for all fractional filling 
factors and peaks around $\nu = 1/2$. 
Unfortunately, finite-size effects do not make it possible 
for us to obtain conclusive results for arbitrary $\nu$, 
except for simple cases such as $\nu = 1/3$. 
This is because the hierarchy FQH states have more complicated edge structure,
which necessarily leads to stronger finite-size effects; the situation is even
worse for compressible bulk like that that of $\nu = 1/2$, because the gapless
bulk and edge excitations are inevitably mixed together. Here we discuss
another relatively simpler case, $\nu = 2/3$, where the QH liquids can
be understood as a $\nu = 1/3$ hole
condensate embedded in a $\nu = 1$ electron 
condensate.~\cite{wen90,macdonald90,johnson91,kane94}
Therefore, two counterpropagating edge modes exist even without the edge
reconstruction. 
We find, not surprisingly, these two modes ($\Delta M = M_{tot} - M_0 =
\pm 1$) in our numerical calculation for $d = 0.1$, as shown in
Figs.~\ref{dipole2}(a) and (c).  
One mode ($\Delta M = 1$, marked by a solid arrow) 
has a squared dipole matrix element 
almost two orders of magnitude larger than the other 
($\Delta M = -1$, marked by a dotted arrow), 
so the low-temperature spectral function is dominated by a single
mode (the edge magnetoplasmon mode). 
On the other hand, for $d = 2.0$, we find, 
more edge modes which include two $\Delta M = -1$ 
low-energy excitations as shown in Fig.~\ref{dipole2}(d).
These new modes can roughly be regarded as the results of the 
edge reconstruction of the 
inner $\nu = 1/3$ hole condensate. 
There is no evidence that the outer $\nu = 1$ edge can be 
reconstructed since we model samples with sharp cleaved edges. 
The dipole spectral function, however, is now dominated by one of the 
$\Delta M = -1$ modes [marked by a solid arrow in
Fig.~\ref{dipole2}(c)], with a squared spectral weight more than 
two orders of magnitude greater than those of 
the $\Delta M = 1$ modes (the lowest one marked by a dotted arrow). 
Thus, quite similar to the $\nu = 1/3$ case, we find 
additional modes due to 
edge reconstruction at larger $d$. We also find the transition processes that 
dominate the dipole spectral function and microwave conductivity change due to
edge reconstruction.
Also similar to the $\nu = 1/3$ case, the dipole spectral function becomes
dominated by bulk excitations at high temperatures 
and the low-frequency spectral weight gets suppressed.
Thus the calculation of the dipole spectral function 
at $\nu = 2/3$ also finds additional low-energy modes 
generically arise from the reconstruction of fractional QH edges, and they
make important contribution to the low-frequency dipole spectral function. Such
contribution, however,
get suppressed at high temperatures. Thus this behavior is not specific to
principal bulk filling, in agreement with 
the enhanced microwave conductivity in antidot samples in the entire fractional
filling range, and its suppression at higher temperature.


\section{Single electron spectral function and edge tunneling}
\label{sec:tunnel}

Numerical calculations of single electron spectral function at the edge
in finite size systems have been
performed by Palacios and MacDonald.~\cite{palacios96}
They considered a QH droplet with Coulomb interaction, but
without a physically realistic confining potential. 
In particular, they calculated the squared matrix elements between the
ground state of an $N$-electron system to the low-lying states of the
corresponding ($N+1$)-electron system at $\nu = 1/3$. 
These numerical results can be compared to those obtained by the CLL
theory, which predicts that the low-lying energy spectrum of a QH
system at principal filling fraction, such as $\nu = 1/3$,  
can be described by a branch of single-boson edge states with 
angular momentum $l$ ($l=$ 1,2,3, ...) and energy $\epsilon_l$.
In the CLL language, we can label each low-energy state by a set of 
occupation numbers $\{n_l\}$, whose total angular momentum and 
energy are $M = M_0 + \Delta M = M_0 + \sum_l l n_l$ and 
$E = E_0 + \Delta E = E_0 + \sum_l n_l \epsilon_l$, respectively, 
where, $M_0$ and $E_0$ are total angular momentum and energy
of the corresponding ground state. 
Palacios and MacDonald~\cite{palacios96} found excellent 
agreement between the squared matrix elements,
$T(\{n_l\}) = | \langle \psi_{\{n_l\}} (N+1) | 
c^{\dagger}_{3N + \Delta M} | \psi_0 (N) \rangle |^2$,
calculated numerically in the microscopic model and 
those calculated based on the CLL theory. 
Note that $M_0 (N+1) - M_0 (N) = 3N$ is the difference in total angular
momenta between the $N$- and ($N+1$)-electron ground states. 
Such a comparison is made possible by the unambiguous identification 
of the low energy spectrum in terms of $\{n_l\}$ (based on $\Delta M$
and $\Delta E$), as well as in the agreement of the 
corresponding values of $T(\{n_l\})$,
at least for $M \le 4$.

In this section, we study the single electron spectral function in the
presence of a physically realistic edge confining potential, 
generated by a layer of background charge, distributed uniformly 
on a disk at distance $d$ above the 2DEG. 
As discussed earlier, the single electron spectral function is directly 
measured in edge tunneling experiments.
We are interested in finding out how the electron spectral function 
is affected by the presence of the edge confining potential, 
which mixes the eigenstates in the system without the confinement.
In particular, we are interested in whether and how the electron spectral
function is affected by edge reconstruction.  

\begin{figure}
{\centering \includegraphics{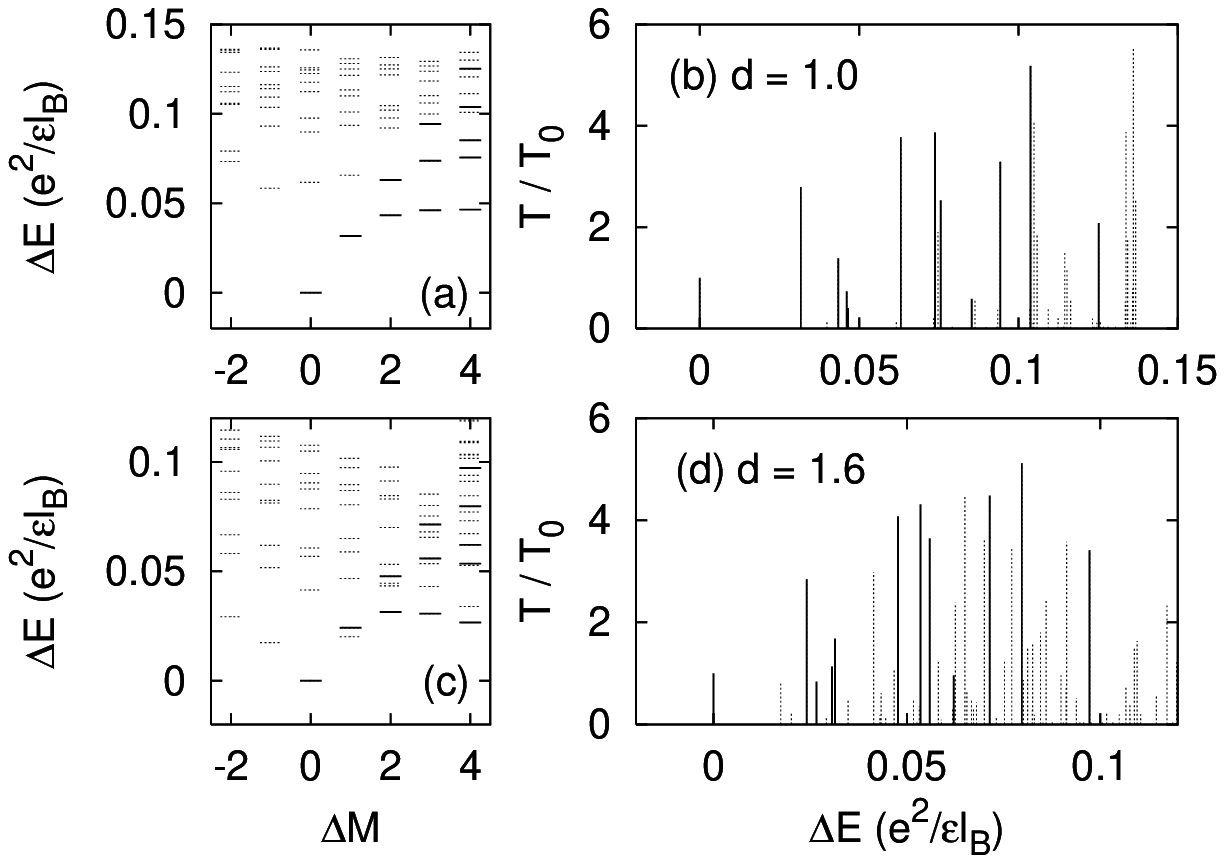} }
\caption{
\label{tunnel}
(a) Spectrum and (b) normalized spectral weights, $T(\{n_l\})/T_0$,
for tunneling one additional electron into a 6-electron system 
at $\nu = 1/3$ for $d = 1.0 l_B$. 
(c) Spectrum and (d) normalized spectral weights, $T(\{n_l\})/T_0$,
for tunneling one additional electron into a 6-electron system 
at $\nu = 1/3$ for $d = 1.6 l_B$. 
In all figures, solid lines repreent states that have significant contribution 
to the electron spectral function, whose  
corresponding matrix elements are listed in Table~\ref{sme}. 
}
\end{figure}

We calculated the tunneling spectral weights for $\nu = 1/3$ by
adding one more electron into a system of 6 electrons.  
The realistic edge confining potential 
is generated by the appropriate background charge, 
which neutralizes the resulting system. 
For $d = 1.0$ (before reconstruction), we plot the spectrum 
and electron spectral weights in Fig.~\ref{tunnel}(a) and (b), respectively. 
As in the absence of the confining potential, we can identify 
the lowest edge excitations for each $M = M_0 + \Delta M$ subspace 
in the low-energy excitation spectrum as 
the single-boson edge state with $n_l = \delta_{l,\Delta M}$ (with
excitation energy $\epsilon_l$). 
Thus, the family of edge states can be unambiguously identified,
since their $\Delta M$ and $\Delta E$ must be simultaneously written 
as linear combinations of $l$ and $\epsilon_l$, respectively. 
We list the tunneling spectral weights for $\Delta M \le 4$ 
in Table~\ref{sme}, 
which are highlighted by solid lines in Fig.~\ref{tunnel}. 
These matrix elements are rather close to those obtained in the absence of the
confining potential,~\cite{palacios96} and consistent with the predictions 
of the CLL theory (for infinite system).

For reconstructed edges on the other hand, the situation becomes more
complicated since there are additional nonchiral boson excitations, all
of which are coupled in general. 
However, it has been proposed~\cite{chamon94} that 
in the strong coupling limit 
(for the unscreened long-range Coulomb interaction which has logarithmic 
singularity in the long-wave length limit)
one mode, which represents
the total charge density mode, may dominate and behaves just like the
single branch below the edge reconstruction transition.
For $d = 1.6$ (after reconstruction), we plot the tunneling spectrum 
and spectral weights in Fig.~\ref{tunnel}(c) and (d), respectively. 
For the lowest two states with $\Delta M = +1$, we find that 
the squared tunneling matrix element for the lower state is 0.213 
and for the second lowest state is 2.845; this is close to 3
which is predicted for the $\{1000\}$ mode in the CLL theory. 
For this reason, we may identify the two states as members of the neutral mode 
and the charge mode, respectively (also the charge mode is indeed expected to
have higher energy). 
In addition, we find that the corresponding squared matrix elements 
for the ground states in $\Delta M =$ 2-4 subspaces are 
1.681, 1.320, and 0.839, respectively. 
These numbers are close to 1.5, 1, and 0.75 
predicted for the edge modes in the corresponding subspaces.
So these three states may be identified as $\{0100\}$, $\{0010\}$, and 
$\{0001\}$ of the charge mode. 
If the distinction of charge and neutral modes can be made 
in this way and 
the charge mode indeed controls the tunneling behavior, 
we should be able to generate a family of excitations 
with angular momenta and energies which can be calculated 
according to the four charge excitations identified so far. 
We should also be able to find, near the calculated energies 
in the corresponding angular momentum subspace, excitations 
with squared matrix elements close to the predictions 
in the CLL theory. 
This indeed seems to be the case, as shown in Table~\ref{sme}. 
The listed states are, again, highlighted in Fig.~\ref{tunnel}. 
We emphasize that the squared matrix elements for  
the rest of the low-lying states ($\Delta E < 0.05 e^2/\epsilon l_B$)
with $\Delta M \neq 0$ are typically very small, 
with the largest one being roughly unity 
(or the ground state-to-ground state value). 
We point out that in order to calculate the matrix elements 
for up to $\Delta M = +4$, we use a smooth edge with $m_{max} = 23$. 
We found that reducing $m_{max}$ (giving a sharper edge) 
has no significant effects on the squared matrix elements 
in the reduced subspaces.
Our results thus suggest that the effect of edge reconstruction on the 
structure of the single electron spectral function is fairly weak; this 
is consistent with the experimental finding that the tunneling exponent is
close to the prediction of the CLL theory at $\nu=1/3$, 
despite the fact that the edges are expected to be
reconstructed.

\begin{table}
\begin{center}
\begin{tabular}{ccc@{\hspace{0.15in}}cc@{\hspace{0.15in}}c@{\hspace{0.15in}}ccc}
\hline \hline
 & & \multicolumn{3}{c}{d = 1.0} & \multicolumn{3}{c}{d = 1.6} & \\
 & & \multicolumn{3}{c}{(no reconstruction)} & \multicolumn{3}{c}{(with
 reconstruction)} & \\
$\Delta M$ & \{$n_l$\} & & $\Delta E$ & $T({\{n_l}\})$ 
& & $\Delta E$ & $T({\{n_l}\})$ & CLL theory \\ \hline
0 & \{0000\} & & 0.0000 & 1.000 & & 0.0000 & 1.000 & 1     \\ \hline
1 & \{1000\} & & 0.0317 & 2.791 & & 0.0241 & 2.845 & 3     \\ \hline
2 & \{2000\} & & 0.0631 & 3.772 & & 0.0477 & 4.074 & 4.5   \\ 
  & \{0100\} & & 0.0434 & 1.383 & & 0.0314 & 1.681 & 1.5   \\ \hline
3 & \{3000\} & & 0.0943 & 3.288 & & 0.0714 & 4.480 & 4.5   \\ 
  & \{1100\} & & 0.0738 & 3.863 & & 0.0559 & 3.644 & 4.5   \\ 
  & \{0010\} & & 0.0461 & 0.734 & & 0.0306 & 1.320 & 1     \\ \hline
4 & \{4000\} & & 0.1252 & 2.083 & & 0.0972 & 3.413 & 3.375 \\ 
  & \{2100\} & & 0.1038 & 5.182 & & 0.0797 & 5.113 & 6.75  \\ 
  & \{1010\} & & 0.0756 & 2.529 & & 0.0535 & 4.312 & 3     \\ 
  & \{0200\} & & 0.0853 & 0.587 & & 0.0621 & 0.955 & 1.125 \\ 
  & \{0001\} & & 0.0465 & 0.402 & & 0.0266 & 0.839 & 0.75  \\ 
\hline \hline
\end{tabular}
\end{center}
\caption{
\label{sme}
Tunneling spectral weights
$T(\{n_l\}) = | \langle \psi_{\{n_l\}} (N+1) | 
c^{\dagger}_{3N + \Delta M} | \psi_0 (N) \rangle |^2$,
for microscopic model at $\nu = 1/3$, 
before ($d = 1.0 l_B$) and after ($d = 1.6 l_B$) 
edge reconstruction transition (normalized to the ground state-to-ground
state matrix element, $T_0$), and for CLL theory. 
The microscopic system contains $N = 6$ electrons 
with corresponding background charge
before an additional electron tunnels into the electron layer.
$\Delta M = M - M_0$ and $\Delta E = E - E_0$ 
(in units of $e^2/\epsilon l_B$)
are total angular momentum and energy 
measured from the corresponding ground state values 
($M_0$ and $E_0$) for the resulting system.
$\{n_l\}$ is a set of occupation numbers of the bosonic 
edge-wave with angular momentum $l$ and energy $\epsilon_l$. 
}\end{table}


\section{conclusions}
\label{sec:conclusion}

In this paper, we have performed exact diagonalization studies on
microscopic models of fractional QH liquids in systems with disk geometry,
and investigated the interplay between electron-electron interaction 
and confining potential due to background charge near the edge. 
We have shown that the edges of fractional QH liquids  
reconstruct when the background charge (dopant) layer is separated far enough
from the electron layer, and the critical distance for this to happen is of 
order one magnetic length. 
The edge reconstruction happens because the electrostatic energy gained 
by moving electrons outward near the sample edge increases logarithmically as
the separation increases, and eventually exceeds the loss of electron 
exchange-correlation energy. 
Such behavior is found to have very weak finite-size effects in most cases, 
even for small systems with 6-9 electrons that we used in our studies. 
Our results suggest that edge reconstruction occurs rather generically in
high-quality AlGaAs/GaAs samples used in experimental studies, 
as the corresponding distance in these samples are typically 
of the order of ten magnetic lengths or larger.

In our studies we have used different types of boundary conditions for 
electronic wave functions near the edge, corresponding to different types of
samples ({\it e.g.}, whether edge is cleaved or not). 
We have demonstrated that the edge reconstruction phenomenon is not 
sensitive to the choices of specific boundary conditions qualitatively, 
be it hard-wall confinement or one that
leads to a smooth confining potential. While different boundary conditions 
lead to quantitatively different critical spacing between dopant and electron
layers, our conclusion that real samples are all in the reconstructed regime
is robust.  

With reconstructed edges, fractional QH liquids can have additional edge
modes that propagate along both directions. 
In general, we find these modes tend to have much lower energy scales than the
edge modes in the absence of edge reconstruction. Therefore, they can have
very important effects on the
low-energy behavior of edge transport and tunneling experiments. 
We have performed calculations on the electric dipole spectral function as well
as single electron spectral function, for systems with and without edge
reconstruction. We find that edge reconstruction affects the dipole spectral
function rather strongly, and its frequency as well as temperature 
dependences compare favorably with microwave conductivity measurements
performed 
in samples with an array of antidots (and their associated edges). On the other
hand we find the electron spectral function at $\nu=1/3$ is not modified 
strongly by edge reconstruction; this is consistent with tunneling experiments 
which find the tunneling exponent at $\nu=1/3$ quantitatively 
close to the prediction of the chiral Luttinger liquid theory, 
despite the presence of edge reconstruction. 


\acknowledgments

We have benefited from a very informative discussion with Matt Grayson on 
Landau level mixing near a sharp cleaved edge.
We also thank Juan Jose Palacios for helpful correspondences on 
the calculation of single electron spectral function. 
This work was supported by NSF grants DMR-9971541 and DMR-0225698 (XW and KY), 
DMR-0086191 (EHR), the State of Florida (XW),
and the A. P. Sloan Foundation (KY).



\newpage

\begin{thebibliography}{99}
\bibitem{note} Here we assume the layer thickness is zero. 
Finite layer thickness will be considered later.
\bibitem{cutoffnote} In our numerical studies such a cutoff that is pushed 
some distance away from the edge is necessarily for having a finite dimensional
Hilbert space.
\end{thebibliography}
\end{document}